\documentclass[twocolumn,aps,pra,showpacs,superscriptaddress,tightenlines]{revtex4-1}
\usepackage{amsmath}
\usepackage{amsfonts}
\usepackage{graphicx}
\usepackage{epsfig}
\usepackage{color}
\usepackage[colorlinks,citecolor=blue]{hyperref}

\begin{document}

\title{Nonreciprocal optical solitons in a spinning Kerr resonator}
\author{Baijun Li}
\affiliation{Key Laboratory of Low-Dimensional Quantum Structures
and Quantum Control of Ministry of Education, Department of
Physics and Synergetic Innovation Center for Quantum Effects and
Applications, Hunan Normal University, Changsha 410081, China}
\author{\c{S}ahin.~K.~\"{O}zdemir}
\affiliation{Department of Engineering Science and Mechanics, and Materials Research Institute, Pennsylvania State University, University Park, Pennsylvania 16802, USA}
\author{Xun-Wei Xu}
\affiliation{Key Laboratory of Low-Dimensional Quantum Structures
and Quantum Control of Ministry of Education, Department of
Physics and Synergetic Innovation Center for Quantum Effects and
Applications, Hunan Normal University, Changsha 410081, China}
\author{Lin Zhang}
\affiliation{School of Physics and Information Technology, Shaanxi Normal University, Xi'an 710061, China}
\author{Le-Man Kuang}\email{lmkuang@hunnu.edu.cn}
\affiliation{Key Laboratory of Low-Dimensional Quantum Structures
and Quantum Control of Ministry of Education, Department of
Physics and Synergetic Innovation Center for Quantum Effects and
Applications, Hunan Normal University, Changsha 410081, China}
\author{Hui Jing}\email{jinghui73@foxmail.com}
\affiliation{Key Laboratory of Low-Dimensional Quantum Structures
and Quantum Control of Ministry of Education, Department of
Physics and Synergetic Innovation Center for Quantum Effects and
Applications, Hunan Normal University, Changsha 410081, China}

\begin{abstract}
We propose a spinning nonlinear resonator as an experimentally accessible platform to achieve nonreciprocal control of optical solitons. Nonreciprocity here results from the relativistic Sagnac-Fizeau optical drag effect, which is different for pump fields propagating in the spinning direction or in the direction opposite to it. We show that in a spinning Kerr resonator, different soliton states appear for the input fields in different directions. These nonreciprocal solitons are more stable against losses induced by inter-modal coupling between clockwise and counterclockwise modes of the resonator. Our work builds a bridge between nonreciprocal physics and soliton science, providing a promising route towards achieving soliton-wave optical isolators and one-way soliton communications.
\end{abstract}


%
%
%
\maketitle
%

\section{Introduction} \label{Int}

 Solitons are stable waveforms preserving their shape or energy during propagation. They widely exist in natural and artificial systems~\cite{Hasegawa95Solitons,Haus96Solitons,Zabusky65Interaction,Mollenauer80Experimental,Strecker02Formation,Ackemann09Fundamentals}, and are indispensable in applications such as broadband spectroscopy, telecommunications, astronomy and low noise microwave generation. In recent years, rapid advances have been witnessed in creating and utilizing dissipative Kerr solitons in optical microresonators~\cite{Herr14Temporal,Yi15Soliton,Brasch16Photonic,Suh16Microresonator,Yang17Counter,Obrzud17Temporal,Yi17Single,Fujii17Effect,Suh18Gigahertz,Weng19 Polychromatic,Karpov19Dynamics,Shu20A,Wan20Frequency,Fan20lSoliton,Shen20Integrated,Runge20The,Liu20Photonic,Liu20Monolithic,Weng20Gain,Englebert21Parametrically,Bao21Quantum}. Very recently, such solitons have been discovered in a high-Q multimode photonic dimer (pair of strongly-coupled, almost identical nonlinear resonators)~\cite{Xue19Super,Tikan20Emergent,Xu20Spontaneous,Komagata21Komagata,Helgason21Dissipative}, revealing a pleiad of emergent phenomena including gear solitons~\cite{Tikan20Emergent} and solitons operated at an exceptional point~\cite{Komagata21Komagata}. These solitons feature a delicate balance between not only dispersion and nonlinearity but also loss and gain~\cite{Kippenberg18Dissipative}.
 Pioneering works in these fields have provided an attractive platform to build compact and low-power frequency combs that exhibit femtosecond pulses at tens of gigahertz repetition rates~\cite{Kippenberg18Dissipative,Gaeta19Photonic,Fortier1920}, which may enable a wide range of key technologies such as chip-scale clocks, integrated frequency synthesizer~\cite{Spencer18An}, and ultrafast laser ranging~\cite{Zhu18Dual,Trocha18Ultrafast,Suh18Soliton,Riemensberger20Massively,Wang20Long}.

\begin{figure}[hpbt]
\centerline{\includegraphics[width=0.47\textwidth]{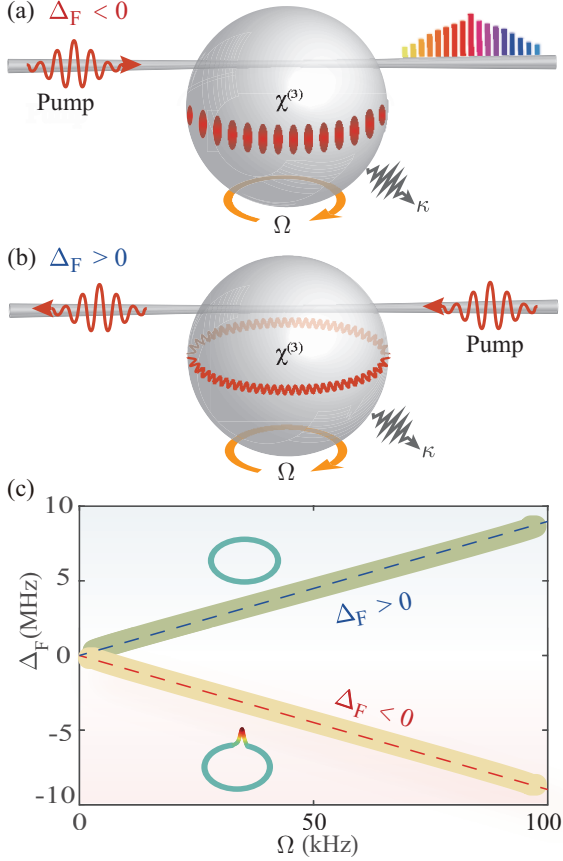}}
\caption{\textbf{Schematic illustration of nonreciprocal solitons.} Different Sagnac frequency shifts $\Delta_\mathrm{F}$ are induced for countercirculating modes in a spinning Kerr microresonator. (a,b) Solitons appear for the input laser in one direction but not the other. (c) The Sagnac-Fizeau shift $\Delta_\mathrm{F}$ versus the angular velocity $\Omega$. Increasing the angular velocity results in a linear opposite frequency shifts for the counterpropagating modes, leading to the formation of nonreciprocal solitons.}
\label{fig:MD}
\end{figure}

In parallel, nonreciprocal optical devices, allowing one-way flow of information and thus playing a key role in backaction-immune communications, have been achieved in nonlinear
 systems~\cite{Shen16Experimental,Shi15Limitations,Fan11An,Xia18Cavity,Scheucher16Quantum,
 Malz18Quantum,Xu20Nonreciprocity,Lai20Nonreciprocal,Sounas17Non-reciprocal}, non-Hermitian
 systems~\cite{Peng14Parity,Chang14Parity,Arkhipov19Scully}, and spinning resonators~\cite{Maayani18Flying,Lu17Optomechanically}. Here we study nonreciprocal control of solitons, which is the first step towards achieving soliton isolators or routers~\cite{Yang20Inverse} and chiral soliton combs~\cite{Tang20Gain}. In contrast to recent works on chiral soliton control using optical gain or dual pumps~\cite{Yang17Counter,Fan20lSoliton,Tang20Gain}, our scheme utilizes another degree of freedom, i.e., the mechanical spinning of the resonator~\cite{Maayani18Flying}. We note that very recently, mechanical rotation has been used in diverse fields such as sound circulation~\cite{Fleury14Sound}, thermal control~\cite{Qiang20Tunable}, one-way quantum optical engineering~\cite{Jiang18Nonreciprocal,Huang18Nonreciprocal,Li19Nonreciprocal,Jiao20Nonreciprocal} and nanoparticle sensing~\cite{Jing18Nanoparticle,Ahn20Ultrasensitive,Zhang20Anti}. In this work, we find that in a spinning nonlinear resonator, nonreciprocal solitons can emerge, {and exist stably even in the presence of strong inter-modal coupling between clockwise (CW) and counter-clockwise (CCW) modes.} Our work is also well complementary to other methods of engineering chiral solitons~\cite{Tang20Gain,Fan20lSoliton}. In fact, by combining optical and mechanical techniques, even more flexible one-way control of solitons can be expected, with practical applications in making and utilizing nonreciprocal soliton devices~\cite{Haus96Solitons,Corcoran20Ultra,Parriaux20Electro,Diddams20Optical}.

\section{Solitons in a spinning resonator}
{The spinning resonator used as the platform in this study can be fabricated by melting the end of a silica glass cylinder and then mounting it on a turbine, as shown in a recent experiment~\cite{Maayani18Flying}. A tapered fiber fabricated from a standard single-mode telecommunications fibre through heat-and-pull method is positioned near this rotating sphere to couple light evanescently into or out of the resonator. Because of the Fizeau drag, the optical paths of counterpropagating lights in the resonator are different, resulting in an irreversible refractive index for the CW and CCW modes. As a result, a $99.6\%$ optical isolation can be realized with such a device at the spinning frequency $\Omega=6.6\mathrm{kHz}$~\cite{Maayani18Flying}.} For such a spinning device, the rapid rotation drags air into the region between the resonator and the fiber, forming a thin layer of air, which helps to stabilize the tapered fiber at a height of several nanometers above the resonator. When there is external disturbance that causes the taper to rise above the stable equilibrium height, the taper will float back and self-adjust itself to its original position~\cite{Maayani18Flying}. The air pressure acting on the fiber is $\Delta T_{air}=(\rho\Delta\theta)T_{air}/L$ which leads to a tiny displacement $d$, where $\rho(\theta)$ indicates the radius (angle) of the winding shape of the deformed region of the fiber, and the whole air pressure on the fiber is~\cite{Maayani18Flying}
\begin{align}
T_{air}=6.19\mu R^{5/2}\Omega\int_{0}^{r}\left(h-\sqrt{r^2-x^2}+r\right)^{-3/2}dx,
\end{align}
where $\mu$ is the viscosity of air, $r (R)$ is the radius of the taper (resonator), and $h=h_0 + d$ represents the taper-resonator separation, with the stationary gap between the fiber and the stationary sphere, $h_0$. The tension on this infinitesimal cylinder due to the local deformation of the fiber is estimated as
\begin{align}
\Delta T_{en}=2F\mathrm{sin}\left(\Delta \theta/2\right)\approx F\Delta \theta,
\end{align}
where $F$ represents the elastic force on the taper and obeys the Hooke law, $\sigma=\epsilon E$. The $\sigma=F/(\pi r^2)$ is the uniaxial stress, $E$ is the Young modulus of silica, and $\epsilon=\delta_L/L$ is the strain with $\delta_L=L^{'}-L$ denotes the length variation between the original length $L$ and the deformation region $L^{'}$. In the case of stable equilibrium, i.e., $\Delta T_{air}=\Delta T_{en}$, we obtain
\begin{align}
T_{air}=2\pi r^2 E \left[\mathrm{arcsin}(\phi)-\phi \right]\approx\pi r^2\phi^3/3,
\end{align}
where $\phi=4Ld/(L^2+4d^2)$, and we have used the approximation for $|\phi|\ll 1$, $\mathrm{arcsin}(\phi)=\phi+\phi^3/6+\cdots$. Therefore, the displacement $d$ can be estimated as
\begin{align}
d=\frac{L}{2}\left(\tau-\sqrt{\tau^2-1}\right)/2,
\end{align}
where $\tau=[\pi r^2 E/(3T_{air})]^{1/3}$. Then, the strain of the taper can be rewritten as $\epsilon=\mathrm{arcsin}(\phi)/\phi-1\approx \phi^2/6$. We find that the strain (i.e., the elastic force) is positively associated with the taper-resonator separation:
\begin{align} \label{S1}
\frac{\partial F}{\partial h}=\pi r^2 E\left(\frac{\partial \epsilon}{\partial d}\right)=\frac{16\pi r^2 E L^2 d(L^2-4d^2)}{3(L^2+4d^2)}>0,
\end{align}
i.e., the elastic force becomes stronger when the air gap gets larger than the stable-equilibrium distance. As a result, the taper is dragged back to its equilibrium position, enabling separation of the taper from the spinning resonator and critical coupling of light into the cavity. We remark that even if the taper is pushed towards the rotating resonator, the taper will not contact or stick to the resonator, which is opposite to the case observed for a static resonator (i.e., the taper may stick to the resonator through van der Waals forces and thus needs to be pulled back to break the connection).

\begin{figure*}[t]
\centerline{\includegraphics[width=0.95\textwidth]{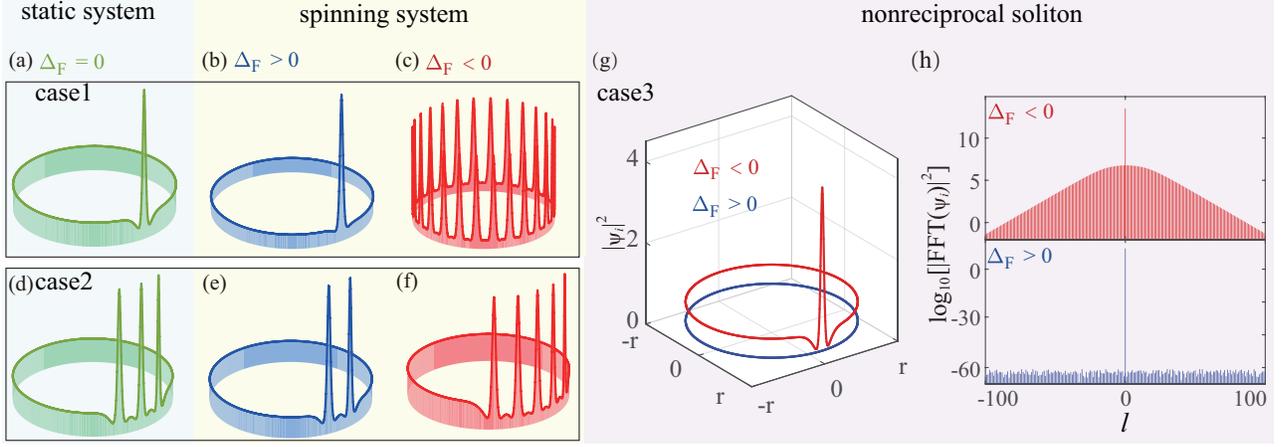}}
\caption{\textbf{Optical solitons with different numbers in different directions.} (a-g) Three-dimensional representations of optical field energy evolution of the soliton mode $|\psi_i|^2$ in the azimuthal direction at $\tau=200$ and (h) corresponding Kerr combs in a static and a spinning Kerr microresonator without backscattering. For $\Delta_{\mathrm{F}}=0$, the solitons emerge with same numbers in different driving directions. For $\Delta_{\mathrm{F}}\ne0$, the detuning $\Delta_{\mathrm{p}}$ for the countercirculating modes are modified by the opposite Sagnac shifts, resulting in solitons  with different numbers in different directions.  {Here, we have shown three different types of nonreciprocal solitons in different parameter ranges.} The initial conditions of the soliton optical fields are Gaussian pulses: (a-c,g-h) $\psi_{0}=0.5+\mathrm{exp}[-(\theta/0.1)^2]$, (d-f) $\psi_{0}=0.5+\mathrm{exp}[-(\theta/0.55)^2]$, the rotation velocity (b,c) $\Omega=2.0~\mathrm{kHz}$, (e,f) $\Omega=350~\mathrm{Hz}$, (g,h) $\Omega=2.2~\mathrm{kHz}$, the dimensionless detuning (a-f) $\Delta_ \mathrm{p}=2$, (g,h) $\Delta_\mathrm{p}=2.3$. We used normalized dispersion $\beta=-0.004$, scaled pump intensity $F_i=1.37$ in the simulations. The figures don't display the $xyz$ coordinate axis for convenience since the numbers of solitons can be seen from the waveform.}
\label{fig:ru1}
\end{figure*}

After confirming the mechanical and optical stabilities of the spinning device, we consider the driving laser with frequency $\omega_d$ from the left or the right via evanescent field of the optical fiber~\cite{Spillane03Ideality} (see Fig.~\ref{fig:MD}). The CW and CCW modes of the resonator are coupled to each other via the backscattering process, that may be induced due to scattering centers or material inhomogeneity formed during fabrication. In the spinning resonator, the frequency of incident laser changes slightly, i.e.,
$\omega_{\pm}-\omega=\pm\Omega R n\omega /c,$
where the subscripts $\pm$ denote the light propagating against or along the spinning direction, $c$ is the speed of light in vacuum, and $n$ and $R$ are the refractive index and the radius of the resonator, respectively. Since
$n(\omega_{\pm})\approx n(\omega)+[n(\omega_{\pm})-\omega]\alpha$ and $\alpha=dn(\omega)/d\omega$,  in view of the Lorentz transformation, the speed of light in the spinning resonator can be written as
$c_{\pm}=({u_{\pm}\pm\Omega R})/({1\pm\Omega Ru_{\pm}/c^{2}})$,
with
\begin{equation}
u_{\pm}=\frac{c}{n(\omega_{\pm})}
\approx\frac{c}{n}\mp\frac{\omega}{n}\frac{dn}{d\omega}\Omega R.
\end{equation}
The optical frequencies of the counter-propagating light fields are then given as {
\begin{align}
\nu_{\pm}=\frac{N c_{\pm}}{2\pi R}\approx\frac{N}{2\pi R}\left[\frac{c}{n}\pm\Omega R\left(1-\frac{1}{n^2}-\frac{\omega}{n}\frac{dn}{d\omega}\right)\right],
\end{align}
with $N=1,2,3,\cdots,$ leading to the Sagnac shifts [see Fig.~\ref{fig:MD}(c)]~\cite{Malykin00The,Maayani18Flying}, i.e.,
\begin{align}
\omega_{0}\rightarrow \omega_{0} \pm \Delta_ \mathrm{F},
\end{align}
with
\begin{equation}
\Delta_\mathrm{F}=2\pi(v_{\pm}-v_{0})\approx\omega_{0}
\frac{n\Omega R}{c}\left(1-\frac{1}{n^{2}}
-\frac{\lambda}{n}\frac{dn}{d\lambda}\right),
\end{equation}
where $v_{0}=N c/(2\pi nR)$} and $\omega_{0}=2\pi c/\lambda$ is the pump frequency for the static resonator. The dispersion term ${\mathrm{d} n}/{\mathrm{d} \lambda}$ that characterizes the relativistic origin of the Sagnac effect is relatively small in typical materials ($\sim1\%$)~\cite{Malykin00The,Maayani18Flying}. The light drag can be further enhanced by dispersion, as demonstrated also in an experiment using a moving microcavity~\cite{Qin20Fast}. Here, for convenience, we consider only the CW rotation of the microresonator. Also we focus on the fundamental modes denoted by an integer wave number $l$ around the eigennumber $l_0$ of the pump. The eigenfrequencies of these modes can then be expanded as~\cite{Herr14Temporal}
\begin{align}\label{fre}
\omega_l=\omega_{0}+\sum_{n=1}^{N}\frac{\xi_{n}}{n!}(l-l_{0})^n,
\end{align}
where $\xi_1=d\omega/dl|_{l=l_0}=c/(nR)$ is the free spectral range and $\xi_2=d^2\omega/dl^2|_{l=l_0}$ is the second-order dispersion coefficient responsible for the asymmetric eigenfrequencies.

\begin{figure*}[t]
\centerline{\includegraphics[width=0.95\textwidth]{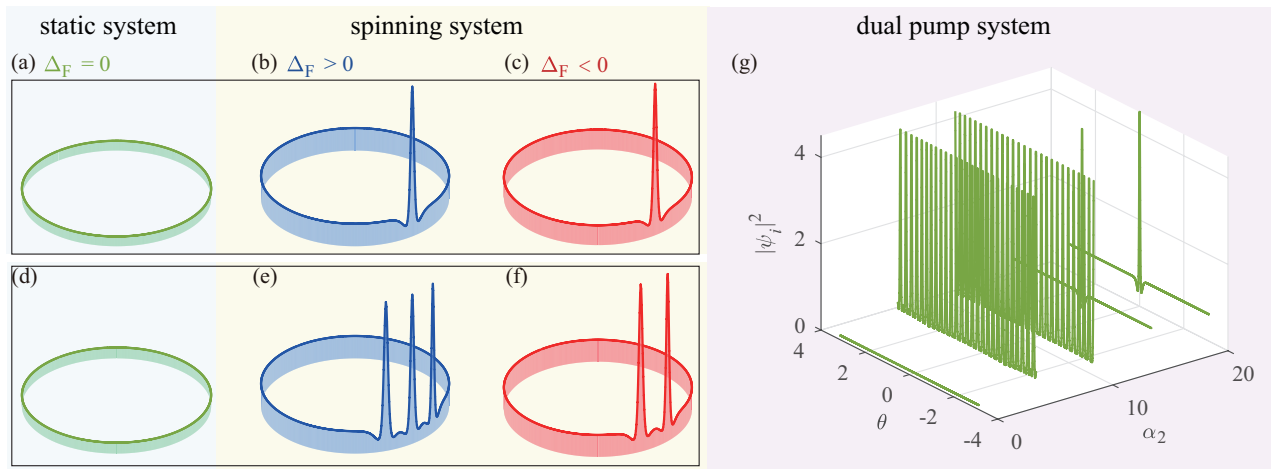}}
\caption{\textbf{Soliton suppression due to random-defect-induced backscattering, and its revival resulting from the rotation-induced compensation.} (a-f) Three-dimensional representations of the optical field energy evolution of the soliton mode $|\psi_i|^2$ in the azimuthal direction at $\tau=200$ in a static and a spinning microresonator when backscattering ($J=1$) is considered. (g) The transient dynamics of the soliton mode $|\psi_i|^2$ at $\tau=200$ as the function of azimuthal angle $\theta$ and the detuning of CCW mode $\alpha_2$ in the dual pump system as reported in Ref.~\cite{Yang17Counter} when backscattering ($J=1$) is considered. For $\Delta_{\mathrm{F}}=0$, solitons are suppressed for increased values of backscattering $J=1$. For $\Omega=60~\mathrm{kHz}$, the soliton {can exist stably.} The initial conditions of the soliton optical fields are Gaussian pulses: (a-c,g) $\psi_{0}=0.5+\mathrm{exp}[-(\theta/0.1)^2]$, (d-f) $\psi_{0}=0.5+\mathrm{exp}[-(\theta/0.55)^2]$, the rotation velocity $\Omega=60~\mathrm{kHz}$, and the normalized detuning (a,d) $\Delta_\mathrm{p}=2.0$, (b,e) $\Delta_\mathrm{p}=-6.9$, (c,f) $\Delta_\mathrm{p}=11$. Other parameters are the same as Fig.~\ref{fig:ru1} and we have set the dimensionless driving strength $F_i=F_j=1.37$ and the scaled detuning of CW mode $\alpha_1=2.2$ in the dual pump system simulations.}
\label{fig:ru3}
\end{figure*}

{Here we consider conventional solitons created in a Kerr resonator, the dynamics of which can be well described by the spatiotemporal Lugiato-Lefever equation~\cite{Lugiato87Spatial,Chembo13Spatiotemporal,Kondratiev20Modulational,Lobanov20Generation,Skryabin20Hierarchy}, with the mode coupling term as used in the experimental works~\cite{Yang17Counter,Yi17Single,Fujii17Effect,Bao21Quantum},}

\begin{align}\label{LLE}
\frac{\partial \psi_{i}}{\partial \tau}=&-[1+i(\Delta_\mathrm{p}-\Delta_\mathrm{sag})]\psi_{i}+i(|\psi_{i}|^2+2|\psi_{j}|^2)\psi_{i}\nonumber\\ &-i\frac{\beta}{2}\frac{\partial^2 \psi_{i}}{\partial \theta^2}+iJ\psi_{j}+F_{i},\nonumber\\
\frac{\partial \psi_{j}}{\partial \tau}=&-[1+i(\Delta_\mathrm{p}+\Delta_\mathrm{sag})]\psi_{j}+i(|\psi_{j}|^2+2|\psi_{i}|)\psi_{j}\nonumber\\ &-i\frac{\beta}{2}\frac{\partial^2 \psi_{j}}{\partial \theta^2}+iJ\psi_{i},
\end{align}
where $\psi_i$ and $\psi_j$ denote {the normalized field amplitudes of the pump and the backscattering, respectively.}  The backscattering process, as observed in the experiments~\cite{Weiss95Splitting,Kippenberg02Modal,Yoshiki15Observation}, is denoted by the strength $J_0$; $\theta\in[-\pi, \pi]$ is the azimuthal angle along the circumference of the resonator, and $\tau=\kappa t/2$ is the dimensionless time. Other scaled parameters are $J=-2J_0/\kappa$, $\Delta_\mathrm{p}=-2\delta/\kappa$, $\beta=-2\xi_2/\kappa$, $\Delta_\mathrm{sag}=2\Delta_\mathrm{F}/\kappa$, where $\kappa=\omega_{0}/Q$ is the optical loss rate, with the quality factor of the resonator given as $Q$, and $\delta=\omega_d-\omega_{0}$ is the detuning between the frequency of the pump laser and the resonance frequency of the resonator. Also, the dimensionless external pump field intensity is found as
\begin{align}
F_i=\sqrt{\frac{4gP}{\hbar\kappa^2\omega_d}}=\sqrt{\frac{4Pn_2c\omega_d}{\kappa^2 n^2V_0}},
\end{align}
where $P$ is the laser power and $g$ is the Kerr coefficient usually given by
\begin{align}
g=\frac{n_2c\hbar \omega_d^2}{n^2V_0},
\end{align}
with $n$ and $n_2$ denoting the linear and nonlinear refraction index of the bulk material, respectively, and $V_0$ representing the effective mode volume of the microresonator. In the following, the cross-phase modulation between counterpropagating modes is neglected since for short pulses it is three order lower than the strong linear coupling, as confirmed experimentally in Ref.~\cite{Fujii17Effect}.

We first consider the case without any backscattering. In our calculations, we use the experimentally feasible parameter values: $\lambda=1550~\mathrm{nm}$, $Q=10^9$~\cite{Pavlov17Soliton}, $R=30~\mathrm{um}$, $\beta=-0.004$, $n=1.44$, and $\Omega=2.2~\mathrm{kHz}$. In Fig.~\ref{fig:ru1}, we plot the intracavity field energy of the soliton mode $|\psi_i|^2$ as a function of the azimuthal angle $\theta$ at $\tau=200$ and corresponding Kerr combs in a static and a spinning microresonator. For a static resonator, our results reproduce and confirm the results of previous studies that have shown the emergence of optical solitons with different numbers under different initial conditions in a reciprocal way [see Figs.~\ref{fig:ru1}(a,d)]~\cite{Godey14Stability}: For a given pump light, the same soliton state is formed regardless of the input pump direction. In contrast, for a spinning device, the system exhibits interesting nonreciprocity. For example, Figs.~\ref{fig:ru1}(b-c,e-f) show that, driving the device from the left or the right (i.e., in the same or opposite direction of the spinning) can lead to two different cases: Soliton crystal state emerges for the pump in one direction whereas a single soliton state emerges for the pump in the other direction [see Figs.~\ref{fig:ru1}(b,c)]. Interestingly, we find that in such a device, it is possible to realize directional switching of soliton numbers by tuning the system parameters: As shown in Figs.~\ref{fig:ru1}(d-f), in a static resonator, only a three-soliton state appears for both input directions~\cite{Godey14Stability}, whereas in a spinning device (with $\Omega=350~\mathrm{Hz}$), a two-soliton state appears for the input from the left (blue curves) and a five-soliton state appears for the same input from the right side (red curves).

Clear signature of nonreciprocal solitons is also observed in Figs.~\ref{fig:ru1}(g,h). By tuning the system parameters $\Delta_\mathrm{p}$ and $\Omega$ properly, one can obtain a soliton-diode behavior. For example, for $\Delta_\mathrm{p}=2.3$ and $\Omega=2.2~\mathrm{kHz}$, a soliton emerges only when driving the system from the left, but it is blocked when driving from the right. This can be explained by different Sagnac drags $\Delta_{\mathrm{F}}$ for the two counter-circulating modes of the microresonator which makes it impossible to maintain the double balance condition (nonlinearity and anomalous dispersion, as well as gain and dissipation) simultaneously for these two distinct modes.

\section{Backscattering-immune features} \label{Backscattering-immune}

\begin{figure}[t]
\centerline{\includegraphics[width=0.49\textwidth]{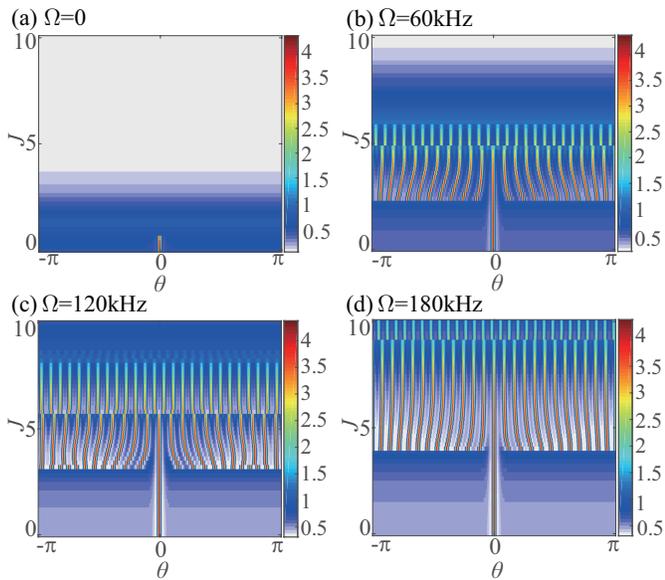}}
\caption{\textbf{The competitive relationship between the backscattering strength and the spinning velocity in nonreciprocal solitons.} The intracavity field energy of the soliton mode $|\psi_i|^2$ as a function of the azimuthal angle $\theta$ at $\tau=200$ and the backscattering strength $J$ with various spinning velocity $\Omega$. The input signal of the optical soliton field is a Gaussian pulse: $\psi_{0}=0.5+\mathrm{exp}[-(\theta/0.1)^2]$, the dimensionless detuning is $\Delta_{\mathrm{p}}=2.1+\Delta_{\mathrm{sag}}$, and the other parameters are the same as Fig.~\ref{fig:ru1}.}\label{fig:ru4}
\end{figure}

Nonreciprocal solitons can also be used to protect optical devices against losses {induced by inter-modal coupling between CW and CCW modes}, without using any specially constructed topological structure or chiral reservoir~\cite{Shalaev19Robust,Yu20Critical}. Figs.~\ref{fig:ru3}(a,d) show that, solitons disappear in a static resonator for increased values of backscattering $J=1$. In contrast, in a spinning device, solitons can exist stably. Figs.~\ref{fig:ru3}(b,e) show that by choosing a spinning velocity, i.e., $\Omega=60\mathrm{kHz}$, {the harmful effect of backscattering can be overcome in a chosen direction. In order to make this advantage clearer, we show in Fig.~\ref{fig:ru4} the competitive relationship between the backscattering strength $J$ and the spinning velocity $\Omega$. We find that as the rotation speed increases, the solitons become more stable even in the presence of a larger backscattering strength.} This suggests that the performance of soliton devices can be further improved by harnessing the power of nonreciprocity, which can be useful in nonreciprocal optical communications or metrology~\cite{Haus96Solitons,Corcoran20Ultra,Parriaux20Electro,Diddams20Optical}.

{The underlying physics can be explained as follows. In a static optical resonator (e.g., microsphere, microring, microtoroid, etc) supporting frequency-degenerate CW and CCW modes, random-defect-induced backscattering may lift the degeneracy by inter-modal coupling and hence result in a symmetric mode splitting~\cite{Weiss95Splitting,Kippenberg02Modal,Yoshiki15Observation}. Various methods have been used to suppress the backscattering by breaking the time-reversal symmetry of the system and thus creating different optical densities in the counterpropagating modes~\cite{Golubentsev84The,MacKintosh88Coherent,Kim19Dynamic,Svela20Coherent}. A recent experiment has demonstrated that~\cite{Maayani18Flying}, spinning the resonator provides a new way to break the time-reversal symmetry. Here, we use spinning-induced nonreciprocity in a nonlinear resonator to achieve one-way flow of solitons. Such spinning-induced nonreciprocal devices can also be used to achieve quantum nonreciprocal effects as studied in Refs.~\cite{Huang18Nonreciprocal,Li19Nonreciprocal,Jiao20Nonreciprocal}.}

For completeness, in Fig.~\ref{fig:ru3} (g), we also show the transient dynamics in the dual pump system as reported in Ref.~\cite{Yang17Counter} when backscattering ($J=1$) is considered. We find that the dual pump scheme can be used to overcome backscattering by choosing a optical detuning $\alpha_2=15$. We emphasize that our scheme, utilizing the mechanical rotation degree of freedom, is well compatible with the other established techniques (e.g., using an optical gain medium or the dual pump technique). In fact, by combining optical and mechanical techniques, even more flexible one-way control of solitons can be expected, with practical applications in making and utilizing backaction-immune soliton devices~\cite{Haus96Solitons,Corcoran20Ultra,Parriaux20Electro,Diddams20Optical}.

\section{Conclusions} \label{C}

In summary, we have proposed a system that uses a spinning Kerr resonator to prepare and engineer nonreciprocal optical solitons. We find that in such a system, different soliton states appear for the input lasers in the CW and CCW directions. These nonreciprocal solitons {are stable against losses induced by inter-modal coupling between CW and CCW modes of the resonator}, which are otherwise detrimental for conventional solitons in static devices. We expect even more flexible control of one-way solitons if mechanical spinning is combined with other well-established optical techniques. In a broader view, our results can stimulate more works on preparing nonreciprocal solitons in systems well beyond spinning photonics, such as in solid devices~\cite{Shen16Experimental}, atoms~\cite{Xia18Cavity}, or synthetic materials~\cite{Sounas17Non-reciprocal,Peng14Parity,Chang14Parity} where nonreciprocity has been demonstrated. {In the future, we will also study nonreciprocal engineering of other kinds of solitons by taking into account the effects of e.g., optical feedback~\cite{Shen20Integrated}, nonideal geometric shape of the resonator, and quantum diffusions~\cite{Bao21Quantum}.}
\bigskip

{\bf Acknowledgements.} We thank Zheng-Yu Wang and Fang-Jie Shu for helpful discussions. H.J. is supported by the National Natural Science Foundation of China (Grants No. 11935006 and No. 11774086) and the Science and Technology Innovation Program of Hunan Province (Grant No. 2020RC4047). L.-M.K. is supported by the NSFC (Grants No. 11935006 and No. 11775075). X.-W.X. is supported by the NSFC (12064010). L.Z. is supported by the NSFC for emergency management project (Grant No.11447025). \c{S}.K.O. acknowledges support from ARO (Grant No.W911NF-18-1-0043), NSF (Grant No. ECCS 1807485), and AFOSR (Grant No. FA9550-18-1-0235). B.J.L. is supported also by Hunan Provincial Innovation Foundation For Postgraduate (No. CX20190339).

\end{document}